\documentclass[preprint,aps,eqsecnum]{revtex4}
\input psfig
\def\be{\begin{equation}}
\def\ee{\end{equation}}
\def\bea{\begin{eqnarray}}
\def\eea{\end{eqnarray}}
\def\ba{\begin{array}}
\def\ea{\end{array}}

\begin{document}
\title{$x-$dependence of the quark distribution functions in the $\chi$CQM$_{{\rm config}}$}
\author{Harleen Dahiya  and Manmohan Gupta}
\affiliation{Department of Physics, Centre of Advanced Study in
Physics, \\ Panjab University, Chandigarh-160 014, India.}
\date{\today}

\begin{abstract}
Chiral constituent quark model with configuration mixing
($\chi$CQM$_{{\rm config}}$) is known to provide a satisfactory
explanation of the ``proton spin problem'' and related issues. In
order to enlarge the scope of $\chi$CQM$_{{\rm config}}$, we have
attempted to phenomenologically incorporate $x-$dependence in the
quark distribution functions. In particular, apart from
calculating valence and sea quark distributions $q_{{\rm val}}(x)$
and $\bar q(x)$, we have carried out a detailed analysis to
estimate the sea quark asymmetries $\bar d(x)-\bar u(x)$, $\bar
d(x)/\bar u(x)$ and $\frac{\bar d(x)-\bar u(x)}{u(x)-d(x)}$ as
well as spin independent structure functions $F_2^p(x)-F_2^n(x)$
and $F_2^n(x)/F_2^p(x)$ as functions of $x$. We are able to
achieve a satisfactory fit for all the above mentioned quantities
simultaneously. The inclusion of effects due to configuration
mixing have also been examined in the case $F_2^p(x)-F_2^n(x)$ and
$F_2^n(x)/F_2^p(x)$ where the valence quark distributions dominate
and it is found that it leads to considerable improvement in the
results. Further, the valence quark structure has also be tested
by extrapolating the predictions of our model in the limit
$x\rightarrow 1$ where data is not available.

\end{abstract}

\maketitle

\section{Introduction}

Chiral constituent quark model ($\chi$CQM), as formulated by
Manohar and Georgi \cite{{manohar}}, has been successful in not
only explaining the ``proton spin crisis'' \cite{emc} but is also
able to account for various general features of the quark flavor
and spin distribution functions including the violation of
Gottfried Sum Rule {\cite{{nmc},{e866},{GSR}}}, baryon magnetic
moments and hyperon $\beta-$decay parameters etc.
{\cite{{eichten},{cheng},{song},{johan}}}. Recently, it has been
shown that configuration mixing, known to be compatible with the
$\chi$CQM (henceforth to be referred as $\chi$CQM$_{{\rm
config}}$), improves the predictions of $\chi$CQM regarding the
spin polarization functions as well as gives an excellent fit to
the baryon magnetic moments \cite{hd}. The successes of
$\chi$CQM$_{{\rm config}}$ have been tested rather well for the
quantities which are $Q^2$ as well as $x$ independent. In this
context, it may be added that in the last few years, there has
been considerable refinement in the measurements of $\bar d-\bar
u$ and $\bar d/\bar u$ asymmetry as well as $\frac{\bar d-\bar
u}{u-d}$ at different values of Bjorken scaling variable $x$. The
NuSea Experiment (E866) \cite{e866} has determined independently
how the sea-antiquark ratio $\bar d/\bar u$ and the difference
$\bar d-\bar u$ vary with $x$ and it has been found that the $Q^2$
dependence is small in these quantities. Earlier, the NMC data was
available for $\bar d-\bar u$ \cite{nmc} for a range of $x$ and
NA51 measurements for $\bar d/\bar u$ \cite{baldit} were
constrained at a single value of $x$. More recently, HERMES has
presented the $x$ dependence of another $u-d$ asymmetry
$\frac{\bar d-\bar u}{u-d}$ \cite{hermes}. This suggests a
corresponding extension of the successes of $\chi$CQM$_{{\rm
config}}$ to quantities which are $x$ dependent but have weak
$Q^2$ dependence.

Similarly, the combinations of proton and neutron structure
functions $F_2^{p,n}(x,Q^2)$, having weak dependence on $Q^2$,
provide a good facility to test the quark distribution functions
at different $x$ values \cite{emc-fpfn,nmc-fpfn}. In particular,
the experimental data is available for the difference of proton
and neutron structure functions $F_2^p(x)-F_2^n(x)$ as well as the
ratio $R^{np}(x)=F_2^n(x)/F_2^p(x)$  \cite{emc-fpfn,nmc-fpfn}. It
has been emphasized in the literature that they provide vital
clues to the valence structure of the nucleons in the limit
$x\rightarrow 1$ thereby having important implications for the
valence quark distribution functions. Also, in this limit, such an
exercise has a bearing on the configuration mixing as well
\cite{{dgg},{Isgur},{yaouanc},mitra}.

It has been emphasized in the neutrino-induced DIS experimental
data from the CDHS experiment that, on the one hand, the valence
quark distributions dominate for $x>0.3$ and it is a relatively
clean region to test the valence structure of the nucleon as well
as to estimate the structure functions and related quantities
\cite{cdhs}. On the other hand, the sea quarks dominate for the
values of $x<0.3$. This becomes particularly interesting for the
$\chi$CQM$_{{\rm config}}$ where the effects of sea quarks and
effects of valence quarks can separately be calculated.  These
latest developments have renewed considerable interest in the
structure of the light-quark sea  as well as valence structure of
the nucleon and suggests a closer scrutiny of the valence and sea
quark distribution functions with an emphasis on their $x$
dependence.

To predict the quark distributions functions at low and
intermediate energy scale, we need to solve the nonperturbative
QCD which is an extremely difficult theoretical problem. The
question of developing a quark model with confining potential
incorporating $x-$dependence in the valence quarks distribution
functions has been discussed in the literature
\cite{Isgur,yaouanc,mitra,bag}. The inclusion of $x-$dependence
however has not yet been successfully derived from first
principles. Instead, they are obtained by fitting parametrizations
to data. It has also been realized in semiphenomenological
analyses \cite{reya} that the light flavor sea quark components
are absolutely necessary to describe the quark distribution
functions. The flavor asymmetry of the sea quark distributions
have been established experimentally \cite{nmc,e866} which seems
definitely non-perturbative and cannot be explained by the sea
quarks radiatively generated through the perturbative QCD.
Therefore, some low-energy nonperturbative mechanism is needed
which generates sea-quark distributions in the nucleon. Recently,
these issues have been discussed in the chiral quark soliton model
\cite{cqsm} which makes a reasonable estimation not only of the
quark distribution functions but also of the antiquark
distributions by taking into account the $x-$dependence in the
distribution functions. More recently, Alwall and Ingelman
\cite{alwall} have derived, in their physical model, the
$x-$dependence in the quark distribution functions from simple
assumptions regarding the nonperturbative properties of the
hadron. Earlier also, similar kind of calculations have been done
by Eichten, Hinchliffe and Quigg in the chiral quark model
\cite{eichten} which describes the sea quark distribution
functions. Isgur \cite{Isgur} has also discussed the
$x-$dependence of the quark distribution functions, induced by
relativistic quenching, in the constituent quark model.

The purpose of the present communication is to phenomenologically
include the  $x-$dependence in the  valence and sea quark
distribution functions in the $\chi$CQM$_{{\rm config}}$ and study
its implications for quark flavor asymmetries dominated by the sea
quark distribution functions as well as spin independent structure
functions dominated by valence quark distribution functions. In
particular, we would like to include $x-$dependence in the
quantities which have weak $Q^2$ dependence, for example, $\bar
d(x)-\bar u(x)$, $\bar d(x)/\bar u(x)$, $\frac{\bar d(x)-\bar
u(x)}{u(x)-d(x)}$, $F_2^p(x)-F_2^n(x)$ and $F_2^n(x)/F_2^p(x)$.
Dependence of the quark distribution functions on $x$ is compared
with the data as well as with the results of several theoretical
models, for example, the chiral quark soliton model \cite{cqsm},
chiral bag model \cite{bag},  a physical model by Alwall and
Ingelman \cite{alwall}, statistical quark model \cite{stat} etc..
Further, it would also be interesting to carry out a detailed
analysis of the implications of configuration mixing in the
quantities $F_2^p(x)-F_2^n(x)$ and $F_2^n(x)/F_2^p(x)$, where the
valence quarks play an important role. Furthermore, the valence
quark structure can also be tested by extrapolating the
predictions of our model in the limit $x\rightarrow 1$ where data
is not available.

\section{Chiral constituent quark model with configuration mixing ($\chi$CQM$_{{\rm
config}}$)}

 The details of $\chi$CQM$_{{\rm config}}$ have already
been discussed in Ref. \cite{hd}, however to facilitate the
discussion as well as for the sake of readability of the
manuscript, some essential details have been presented in the
sequel. The key to understand the ``proton spin problem'', in the
$\chi$CQM formalism \cite{cheng}, is the fluctuation process
\be
  q^{\pm} \rightarrow {\rm GB}
  + q^{' \mp} \rightarrow  (q \bar q^{'})
  +q^{'\mp}\,, \label{basic}
\ee
 where GB represents the Goldstone boson and $q \bar q^{'}  +q^{'}$
 constitute the ``quark sea'' \cite{cheng,song,johan,hd}.
The effective Lagrangian describing interaction between quarks and
a nonet of GBs, consisting of octet and a singlet, can be
expressed as
\be
{\cal L}= g_8 {\bf \bar q}\Phi {\bf q} + g_1{\bf \bar
q}\frac{\eta'}{\sqrt 3}{\bf q}= g_8 {\bf \bar
q}\left(\Phi+\zeta\frac{\eta'}{\sqrt 3}I \right) {\bf q}=g_8 {\bf
\bar q}\left(\Phi'\right) {\bf q}\,, \ee where $\zeta=g_1/g_8$,
$g_1$ and $g_8$ are the coupling constants for the singlet and
octet GBs, respectively, $I$ is the $3\times 3$ identity matrix.
The GB field which includes the octet and the singlet GBs is
written as \bea
 \Phi' = \left( \ba{ccc} \frac{\pi^0}{\sqrt 2}
+\beta\frac{\eta}{\sqrt 6}+\zeta\frac{\eta^{'}}{\sqrt 3} & \pi^+
  & \alpha K^+   \\
\pi^- & -\frac{\pi^0}{\sqrt 2} +\beta \frac{\eta}{\sqrt 6}
+\zeta\frac{\eta^{'}}{\sqrt 3}  &  \alpha K^0  \\
 \alpha K^-  &  \alpha \bar{K}^0  &  -\beta \frac{2\eta}{\sqrt 6}
 +\zeta\frac{\eta^{'}}{\sqrt 3} \ea \right) {\rm and} ~~~~q =\left( \ba{c} u \\ d \\ s \ea
\right)\,. \eea SU(3) symmetry breaking is introduced by
considering $M_s > M_{u,d}$ as well as by considering the masses
of GBs to be nondegenerate
 $(M_{K,\eta} > M_{\pi})$ {\cite{{song},{johan}}}, whereas
  the axial U(1) breaking is introduced by $M_{\eta^{'}} > M_{K,\eta}$
{\cite{{cheng},{song},{johan}}}. The parameter $a(=|g_8|^2$)
denotes the probability of chiral fluctuation  $u(d) \rightarrow
d(u) + \pi^{+(-)}$, whereas $\alpha^2 a$, $\beta^2 a$ and $\zeta^2
a$ respectively denote the probabilities of fluctuations $u(d)
\rightarrow s + K^{-(0)}$, $u(d,s) \rightarrow u(d,s) + \eta$,
 and $u(d,s) \rightarrow u(d,s) + \eta^{'}$.

The most general configuration mixing, generated by the
chromodynamic spin-spin forces \cite{{dgg},{Isgur},{yaouanc}}, in
the case of octet baryons  can be expressed as
\cite{{Isgur},{yaouanc},{full}}
\be
|B \rangle=\left(|56,0^+\rangle_{N=0} \cos \theta +|56,0^+
\rangle_{N=2} \sin \theta \right) \cos \phi +
\left(|70,0^+\rangle_{N=2} \cos \theta^{'} +|70,2^+\rangle_{N=2}
\sin \theta^{'} \right) \sin \phi\,, \label{full mixing} \ee where
$\phi$ represents the $|56\rangle-|70\rangle$ mixing, $\theta$ and
$\theta^{'}$ respectively correspond to the mixing among
$|56,0^+\rangle_{N=0}-|56,0^+ \rangle_{N=2}$ states and
$|70,0^+\rangle_{N=2}-|70,2^+\rangle_{N=2}$ states. For the
present purpose, it is adequate {\cite{hd,{yaouanc},{mgupta1}}} to
consider the mixing only between $|56,0^+ \rangle_{N=0}$ and the
$|70,0^+\rangle_{N=2}$ states and the corresponding ``mixed''
octet of baryons is expressed
\begin{equation}
|B\rangle \equiv \left|8,{\frac{1}{2}}^+ \right> = \cos \phi
|56,0^+\rangle_{N=0} + \sin \phi|70,0^+\rangle_{N=2}\,,
\label{mixed}
\end{equation}
for details of the  spin, isospin and spatial parts of the
wavefunction,  we  refer the reader to reference
{\cite{{yaoubook}}.

\section{Quark distribution functions and related parameters}

The $\chi$CQM$_{{\rm config}}$ incorporates the valence quarks,
the quark sea and the gluons as the effective degrees of freedom.
However, there are no simple or straightforward rules which could
allow incorporation of $x$ dependence in $\chi$CQM$_{{\rm
config}}$.  The kind of parameters we are evaluating in the
present context, at the leading order, involve only the valence
quarks and the sea quarks and their $x-$dependence contributing to
the total quark distribution functions defined as $q(x)=q_{{\rm
val}}(x)+\bar q(x)$ ($q=u,d,s$). Therefore, we would consider the
$x-$dependence of valence quark distributions given as $u_{{\rm
val}}(x)$, $d_{{\rm val}}(x)$, $s_{{\rm val}}(x)$ and the sea
quark distributions given as $\bar u(x)$, $\bar d(x)$, $\bar
s(x)$. To this end, instead of using an {\it ab initio} approach,
we have phenomenologically incorporated the $x-$dependence getting
clues from Eichten {\it et al.} \cite{eichten}, Isgur \cite{Isgur}
and  Le Yaouanc {\it et al.} \cite{yaouanc}. To begin with, we
emphasize the normalization conditions which have to be satisfied
by the quark and antiquarks distribution functions. For example,
\bea \int_0^1 [(u(x)-\bar u(x)]dx&=&\int_0^1 u_{{\rm
val}}(x)dx=2\,, \nonumber\\ \int_0^1 [(d(x)-\bar
d(x)]dx&=&\int_0^1 d_{{\rm val}}(x)dx=1\,, \nonumber\\ \int_0^1
[(s(x)-\bar s(x)]dx&=&\int_0^1 s_{{\rm val}}(x)dx=0\,.
\label{norm}
 \eea Keeping in
 mind the above normalization conditions and following Eichten
 {\it et al.}, we assume the $x-$dependence of the valence and
 sea  quark distribution  functions as
  \bea u_{{\rm val}}(x)&=& 8
(1-x)^3 \cos^2 \phi+4 (1-x)^3\sin^2 \phi+ 8 \sqrt{2} x^4 (1-x)^3
\cos \phi \sin \phi\,, \label{uval} \\ d_{{\rm val}}(x)&=& 4
(1-x)^3 \cos^2 \phi+2 (1-x)^3 \sin^2 \phi- 8 \sqrt{2} x^4 (1-x)^3
\cos \phi \sin \phi\,, \label{dval} \\ \bar u(x)
&=&\frac{a}{12}[(2 \zeta+\beta+1)^2 +20] (1-x)^{10}\,,
\label{baru} \\ \bar d(x) &=&\frac{a}{12}[(2 \zeta+ \beta -1)^2
+32] (1-x)^7\,, \label{bard}
\\
\bar s(x) &=&\frac{a}{3}[(\zeta -\beta)^2 +9 {\alpha}^{2}]
(1-x)^8\,.
 \label{bars}\eea
It should be noted that the effects of configuration mixing have
been incorporated in the valence quark distribution functions
following Le Yaouanc {\it et al.} and Isgur \cite{Isgur}.

After having formulated the $x$ dependence in the valence and sea
quark distribution functions, we now consider the quantities which
are measured at different $x$ and can expressed in terms of the
above mentioned quark distribution functions. As mentioned earlier
also, we are interested only in the quantities which have weak
$Q^2$ dependence because $\chi$CQM does not incorporate $Q^2$
dependence, therefore we will consider only those experimentally
measurable quantities which have no or weak $Q^2$ dependence. To
this end, the most important quantity is the Gottfried integral
which can be expressed in terms of the quark distribution
functions after separating them into valence and sea contributions
and is expressed as \bea I_G &=&\frac{1}{3}\int_0^1 {[u_{{\rm
val}}(x)-d_{{\rm val}}(x)] dx}+\frac{2}{3} \int_0^1 {[\bar
u(x)-\bar d(x)] dx} \nonumber \\ &=&\frac{1}{3}+\frac{2}{3}
\int_0^1 {[\bar u(x)-\bar d(x)] dx}
\,, \label{devgsr} \eea  where we have used the above
normalization conditions. The quantity $\bar u(x)-\bar d(x)$ can
be obtained from the E866 data \cite{e866} at different $x$ values
and calculated in the present context from Eqs. (\ref{baru}) and
(\ref{bard}). Similarly, the $\bar d(x)/\bar u(x)$
{\cite{e866,baldit}} asymmetry measured through the ratio of muon
pair production cross sections $\sigma_{pn}$ and $\sigma_{pp}$,
can expressed in the present case in terms of the sea quark
distribution functions and its variation with $x$ can be studied.
Further, the $x$ dependence of $\frac{\bar d(x)-\bar
u(x)}{u(x)-d(x)}$ \cite{hermes} can be calculated using Eqs.
(\ref{uval}) to (\ref{bard}) as well as the normalization
conditions from Eq. (\ref{norm}).

To study the $Q^2$ independent combinations of proton and neutron
structure functions $F_2^{p,n}(x,Q^2)$, we write the basic spin
independent structure functions in terms of the valence and sea
quark distribution functions as follows
\be
F_1(x,Q^2)=\frac{1}{2}\sum_{q=u,d,s} e_q^2[ q_{{\rm val}}(x)+\bar
q(x)], \label{f1} \ee
\be
F_2(x,Q^2)=2 x F_1(x,Q^2). \label{f2} \ee The difference of proton
and neutron structure functions  $F_2^p(x)-F_2^n(x)$ as well as
the ratio $R^{np}(x)=F_2^n(x)/F_2^p(x)$ which again have weak
$Q^2$ dependence can be obtained using Eqs.
(\ref{uval}-\ref{bars}).

For the strange-quark content of the nucleon, the relevant
observables \cite{ao} measured over a range of $x$ are given as
\bea \rho&=&2\frac{\int_0^1{x \bar
s(x)dx}}{\int_0^1{x(u(x)+d(x))dx}}\,, \label{rho}\\
\kappa&=&2\frac{\int_0^1{x \bar s(x)dx}}{\int_0^1{x(\bar u(x)+\bar
d(x))dx}}\,. \label{kappa}
 \eea
The parameters $\rho$ and $\kappa$ represents the ratio between
the strange and nonstrange quarks of the sea. With the ratio
$\rho/\kappa$ we obtain the ratio between the nonstrange
antiquarks and quarks in the nucleon as \be \frac{\rho}{\kappa}=
\frac{\int_0^1{x(\bar u(x)+\bar
d(x))dx}}{\int_0^1{x(u(x)+d(x))dx}}\ee

\section{Input parameters}
Before calculating the valence and sea quark distribution
functions and the related phenomenological quantities such as
$\bar d(x)-\bar u(x)$, $\bar d(x)/\bar u(x)$, $\frac{\bar
d(x)-\bar u(x)}{u(x)-d(x)}$, $F_2^p(x)-F_2^n(x)$ and
$F_2^n(x)/F_2^p(x)$ to study $x$ dependence, we have to first fix
the $\chi$CQM$_{{\rm config}}$ parameters ($a$, $\alpha$, $\beta$
and $\zeta$ representing the probabilities of fluctuations to
pions, $K$, $\eta$, $\eta^{'}$) coming in the sea quark
distribution functions $\bar u(x)$, $\bar d(x)$ and $\bar s(x)$ as
well as the mixing angle $\phi$ coming in the valence quark
distribution functions $u_{{\rm val}}(x)$ and $d_{{\rm val}}(x)$.
The mixing angle $\phi$ is fixed from the consideration of neutron
charge radius \cite{{yaouanc},{mgupta1}}, for the other parameters
we have used $\Delta u$, $\Delta_3$, $\bar u-\bar d$ and $\bar
u/\bar d$ as inputs. The range of the coupling breaking parameter
$a$ can be easily found by considering the spin polarization
function $\Delta u$, by giving the full variation of parameters
$\alpha$, $\beta$ and $\zeta$ from which one finds $0.10 \lesssim
a \lesssim 0.14$. The range of the parameter $\zeta$ can be found
from  $\bar u/\bar d$ using the latest experimental measurement
\cite{e866} and it comes out to be $-0.70 \lesssim \zeta \lesssim
-0.10$. Using the above found ranges of $a$ and $\zeta$ as well as
the latest measurement of $\bar u-\bar d$ asymmetry \cite{e866},
$\beta$ comes out to be in the range $0.2\lesssim \beta \lesssim
0.7$. Similarly, the range of $\alpha$ can be found by considering
the flavor non-singlet component $\Delta_3$ and it comes out to be
$0.2 \lesssim \alpha \lesssim 0.5$. After finding the ranges of
the coupling breaking parameters, we have carried out a fine
grained analysis by fitting $\Delta u$, $\Delta d $, $\Delta_3$
\cite{PDG} as well as $\bar u-\bar d$, $\bar u/\bar d$ \cite{e866}
leading to the following set of best fit parameters
\be
a=0.13\,,~~~\alpha=\beta=0.45\,,~~~ \zeta=-0.10\,. \ee   To check
the results in $\chi$CQM$_{{\rm config}}$, we have in Table
\ref{parameters}, presented certain parameters related to spin
polarization functions and quark distribution functions having
implications for the above set of $\chi$CQM parameters. The
details have already been discussed in Ref. \cite{zeta}. On
comparing the results for these quantities with data, it is
evident from the table that we are able to obtain an excellent
agreement in the case of $\Delta d$, $\Delta_8$, baryon octet
magnetic moments $\mu_{p}$, $\mu_{n}$, $\mu_{\Sigma^-}$,
$\mu_{\Sigma^+}$, $\mu_{\Xi^o}$ and $\mu_{\Xi^-}$. Similar
agreement is obtained in the case of quark distribution functions.
Therefore, we will be using the above set of $\chi$CQM parameters
for our calculations.

\section{Results and Discussion}

After having incorporated $x$ dependence in the valence and sea
quark distribution functions as well as fixing the
$\chi$CQM$_{{\rm config}}$ parameters, we now discuss all the
$Q^2$ independent quantities at different values of $x$ and
compare them with the available experimental data as well as other
theoretical models. To begin with, in Figs. \ref{udsval} and
\ref{udssea}, we have shown how the valence and sea quark
distributions of the proton vary with the Bjorken scaling variable
$x$. Although no experimental data is available for these
quantities, we have plotted them in order to make a comparison
with other theoretical results at different $x$. A cursory look at
the plots clearly indicates that our results seem to be well in
line with the suggestions of the chiral quark soliton model
\cite{cqsm} as well as a physical model by Alwall and Ingelman
\cite{alwall}. In order to compare our results with their results,
we can compare the magnitude of quark distribution functions at
some particular values of $x$. For example, in Fig. \ref{udsval},
$x u_{{\rm val}}(x)$ has a peak at around $x=0.25$ and the maximum
value goes upto $0.9$. This is in excellent agreement with the
valence quark distribution function in the chiral quark soliton
model \cite{cqsm} as well as a physical model by Alwall and
Ingelman \cite{alwall}. Similarly, we can compare the sea quark
distribution function $x \bar u(x)$. In this case also, we find
that the peak is at around $x=0.1$ and the maximum value goes upto
$0.1$ also in agreement with the above mentioned models. Similar
agreement is found from the comparison of the case of $x d_{{\rm
val}}(x)$ and $x s_{{\rm val}}(x)$ as well as $x \bar d(x)$ and $x
\bar s(x)$.

From these plots, we are also able to describe some general
aspects of the valence and sea quark distribution functions. From
Figs. \ref{udsval} and \ref{udssea}, one can easily find out that
the valence quark distributions predictions vary as
\[u_{{\rm val}}(x)>d_{{\rm val}}(x)>s_{{\rm val}}(x),\]
whereas the sea quark or the antiquark distributions vary as
\[\bar d(x)>\bar u(x)>\bar s(x).\] It would be important to mention here
that the sea quarks do not contribute at higher values of $x$,
therefore in Fig. \ref{udssea}, we have taken the region
$x=0-0.5$. Beyond this $x$ region the contribution  of the sea
quarks is negligible.  A careful study of the plots brings out
several interesting points. It is evident from Fig. \ref{udsval}
that the valence quark distribution is spread over the entire $x$
region and also that there is $u$ quark dominance when
$x\rightarrow 1$. From Fig. \ref{udssea}, the variation of the
magnitudes of $\bar u(x)$, $\bar d(x)$ and $\bar s(x)$ shows that
there is an excess of $\bar d$ quarks over $\bar u$ quarks in the
regime where the sea quarks contribute.  It is also clear from the
plots that as the value of $x$ increases, the sea contributions
decrease and when $x>0.35$, the contributions are completely
dominated by the valence quarks. This point would be elaborated
further while discussing the $\bar d(x)-\bar u(x)$ asymmetry which
is very much in agreement with the NMC \cite{nmc} and the more
recent E866 asymmetry data \cite{e866}. A comparison of the
valence and sea quark distribution functions also brings out that
the contribution of the sea quarks is less than the valence quarks
by approximately 10\%. These observations are in line with the
observations of other models \cite{cqsm,bag,alwall,stat}.

After having discussed some of the general aspects of the
variation of the valence and sea quark distribution functions with
$x$, we now discuss the variation of some of the well known
experimentally measurable quantities. The variation of the
magnitudes of $\bar u(x)$, $\bar d(x)$ and $\bar s(x)$ are of
special interest as it explains the $\bar d(x)-\bar u(x)$ as well
as the $\bar d(x)/\bar u(x)$ asymmetry. A flavor symmetric sea,
$\bar u$=$\bar d$, leads to the Gottfried sum rule
$I_G=\frac{1}{3}$. A measurement of the Gottfried integral  by NMC
\cite{nmc} and E866 \cite{e866} has resulted in $I_G=0.235 \pm
0.026$ and $0.254\pm 0.005$ respectively. If isospin symmetry
holds, a global flavor asymmetry $ \int_0^1 (\bar d(x) -\bar
u(x))dx \approx 0.15$ and $0.12$ would account for the NMC and
E866 result. It is however interesting to note that the E866
experiment has measured the $x$ dependence of $\bar d(x)-\bar
u(x)$ and $\bar d(x)/\bar u(x)$ asymmetries independently and we
would henceforth be considering the experiment data of  the E866
to compare with the results of our model. In Figs. \ref{bdu} and
\ref{dbyu}, the $\chi$CQM results for the $\bar d(x)-\bar u(x)$,
$\bar d(x)/\bar u(x)$ asymmetries are plotted at different $x$
values. A cursory look at the figures shows that the results of
our model calculations show a satisfactory overlap with the
available E866 data. It would be important to mention that these
quantities provide important constrains on a model that attempts
to describe the origins of the nucleon sea and has very important
implications for the role of sea quarks in the low $x$ regions. In
other words, these asymmetries provide a direct determination of
the ``quark sea'' contribution in the nucleon.  It is clear from
the plots that when $x$ is small $\bar d(x)-\bar u(x)$ and $\bar
d(x)/\bar u(x)$ are large implying the dominance of sea quarks in
this region. In fact, the sea quarks dominate only in the region
where $x$ is smaller than 0.35. On the other hand, when $x
\rightarrow 1$, $\bar d-\bar u$ tends to 0 implying that there are
no sea quarks in this region because the $\bar u$ and $\bar d$
quarks are generated from the ``quark sea'' in $\chi$CQM. In this
case also, our results are well in agreement with the other
theoretical results \cite{cqsm,bag,alwall,stat}

In Fig. \ref{herm-asym}, we have plotted the phenomenological
results for the $\frac{\bar d(x)-\bar u(x)}{u(x)-d(x)}$ asymmetry
at different values of $x$ and have compared them with the
available semi-inclusive DIS experimental data \cite{hermes}. The
values of $\frac{\bar d(x)-\bar u(x)}{u(x)-d(x)}$ are positive
over the region $0.02<x<0.35$ again showing an excess of $\bar d$
over $\bar u$. It is also evident from the plot that more
experimental input is required for the quantity to be compared
with the predictions of the model as the error contained in the
data is too large.  Therefore, a refinement in the data would not
only test the $x$ dependence in the quark distribution functions
but also the extent to which the quark sea contributes in the low
$x$ regime.

The above mentioned quantities test the sea quark distributions,
dominant in the low $x$ regime, reasonably well. It would be
important to mention here that the results presented here depend
mainly on the $\chi$CQM parameters. However, to test the inclusion
of $x$ dependence in the high $x$ regime, it becomes essential to
study the valence quark distributions which play an important role
in the proton and neutron structure functions and in this case the
results depend mainly on the assumed parameterization of the
$x-$shape. In Figs. \ref{fpfn} and \ref{rnp}, we have plotted the
phenomenological results for the difference and the ratio of the
structure functions ($F_2^p(x)-F_2^n(x)$ and
$R^{np}=F_2^n(x)/F_2^p(x)$) after the inclusion of configuration
mixing and have compared them with the available data
\cite{nmc-fpfn}. A cursory look at the plots immediately brings
out that our phenomenological results have a very good overlap
with the data. To make a comparison at some particular values of
$x$, in Table \ref{fpfn-table}, we have also presented the results
for $F_2^p(x)-F_2^n(x)$ and $F_2^n(x)/F_2^p(x)$ at different
values of $x$. A careful scrutiny of the plots and the table
reveals several important points. As the data in the case of the
above mentioned structure functions is available for a broader
range of $x$ as compared to the data available for the sea quark
asymmetries therefore, it becomes interesting to highlight the
importance of valence and sea quarks in different $x$ regimes. We
have carried out a detailed analysis for the valence and sea quark
contribution to $F_2^p(x)-F_2^n(x)$ and it is found that in the
low $x$ regime the contributions are dominated by sea quarks
whereas, as we go to higher $x$ values the contributions are
completely dominated by the valence quarks. It is however
interesting to note that the valence and sea quark distributions
contribute in the right direction to give an excellent overall fit
to $F_2^p(x)-F_2^n(x)$.

Similarly in the case of $R^{np}$, the agreement of our
phenomenological results with the data are excellent over the
entire $x$ range. In this case also the valence and sea quark
distributions contribute in the right direction for the entire $x$
range. As there is no data available for $x>0.7$, it therefore
becomes more interesting to predict the values at $x \rightarrow
1$ and determine the valence structure of the nucleon.
Experimentally, $ R^{np}$ decreases monotonically with the Bjorken
scaling variable $x$ from $ R^{np}\simeq 1$ at $x\simeq0$ to $
R^{np}\simeq \frac{1}{3}$ at $x\simeq0.7$. In our model we predict
\bea R^{np}(x=0.05)&=&0.920\,, \nonumber \\ R^{np}(x=1)&=&0.238\,,
\eea apart from the values at different $x$. The results compare
extremelly well with the experimental figures. Our results are
also in line with other theoretical models such as the chiral
quark soliton model \cite{cqsm}, statistical quark model
\cite{stat} as well as a physical model by Alwall and Ingelman
\cite{alwall} particularly when the value of $x$ is small.
However, when we go to higher values of $x$, our model predictions
are in better agreement with data when compared to the above
mentioned theoretical models. This can easily be seen by comparing
the plots of these models with our predictions.

As has been mentioned earlier, the inclusion of configuration
mixing generated by spin-spin forces
\cite{{dgg},{Isgur},{yaouanc}} in the naive constituent quark
model provides a natural explanation for many of the nucleon
properties including the behaviour of the difference and the ratio
of the structure functions $F_2^p(x)-F_2^n(x)$ and
$F_2^n(x)/F_2^p(x)$ as $x\rightarrow 1$ where the valence quarks
play a dominant role. Therefore, to emphasize the importance of
configuration mixing, we have have also carried out the
calculations without configuration mixing and presented them in
Table \ref{fpfn-table}. In order to appreciate the role of $x$
dependence along with configuration mixing, we first compare the
results of the $\chi$CQM with those of the $\chi$CQM$_{{\rm
config}}$. We find that the configuration mixing effects a uniform
improvement in the case of structure functions compared to those
without configuration mixing. We also find that the inclusion of
$x$ in the quark distribution functions predicts the results in
the right direction even when configuration mixing is not
included, however, when configuration mixing is included, these
show considerable further improvement. This indicates that both
$x$ dependence and configuration  mixing are very much needed to
get an overall fit.

To test the validity of the model as well as for the sake of
completeness, in Table \ref{avg-table}, we have presented the
results of our calculations for the quantities whose data is
available over a range of $x$ or at an average value of $x$.  In
these quantities also we find a good overall agreement with the
data. We have already discussed the excellent agreement achieved
in the case of $\bar d(x)-\bar u(x)$ asymmetry at different values
of $x$. The data for this quantity is also available for the
ranges $x=0-1$ and $x=0.05-0.35$ and is given as $\int_0^1{(\bar
d(x)-\bar u(x))dx}=0.118 \pm 0.012$ and $\int_{0.05}^{0.35}{(\bar
d(x)-\bar u(x))dx}=0.0803 \pm 0.011$. We find that, in our model,
we are also able to give an almost perfect fit for the data
available at these $x$ ranges. It also needs to be mentioned that
the data for the strangeness dependent parameters mentioned in
Eqs. (\ref{rho}) and (\ref{kappa})  is also available for the
range $x=0-1$. These quantities lead to some important
observations. As there are no strange quarks in the valence
structure of the nucleon, therefore the valence contribution to
the strangeness parameters $\rho$ and $\kappa$ is zero. This
implies that the contribution is purely from the quark sea.  The
quality of numerical agreement in this case can be assessed only
after the data gets refined.

\section{Summary and Conclusions}
To summarize, in order to enlarge the scope of $\chi$CQM$_{{\rm
config}}$, we have phenomenologically incorporated the
$x-$dependence in the quark distribution functions following  Le
Yaouanc {\it et al.} and Eichten {\it et al.}. After having
calculated valence and sea quark distributions $q_{{\rm val}}(x)$
and $\bar q(x)$, we have carried out a detailed analysis to
estimate $\bar d(x)-\bar u(x)$, $\bar d(x)/\bar u(x)$, $\frac{\bar
d(x)-\bar u(x)}{u(x)-d(x)}$, $F_2^p(x)-F_2^n(x)$ and
$F_2^n(x)/F_2^p(x)$ as functions of $x$ where we are able to
achieve an excellent agreement with data as well as other
theoretical studies for all the quantities simultaneously.
Implications of configuration mixing have also been studied and it
is found that the inclusion of $x$ predicts the results in the
right direction even when configuration mixing is not included,
however, when configuration mixing is included, these show
considerable further improvement. The valence quark structure has
also been tested by extrapolating the predictions of our model in
the region $x\rightarrow 1$ where data is not available. The
strangeness dependent parameters $\rho$ and $\kappa$ coming purely
from the sea degrees of freedom leads to some important
observations regarding the dependence of the sea quark distibution
functions on the Bjorken scale variable $x$ which can perhaps be
substantiated by further refinement in the data.

\section{acknowledgements}
H.D. would like to thank DST (Fast Track Scheme), Government of
India, for financial support and the chairman, Department of
Physics, for providing facilities to work in the department.

\begin{figure}
  \centerline{\psfig{figure=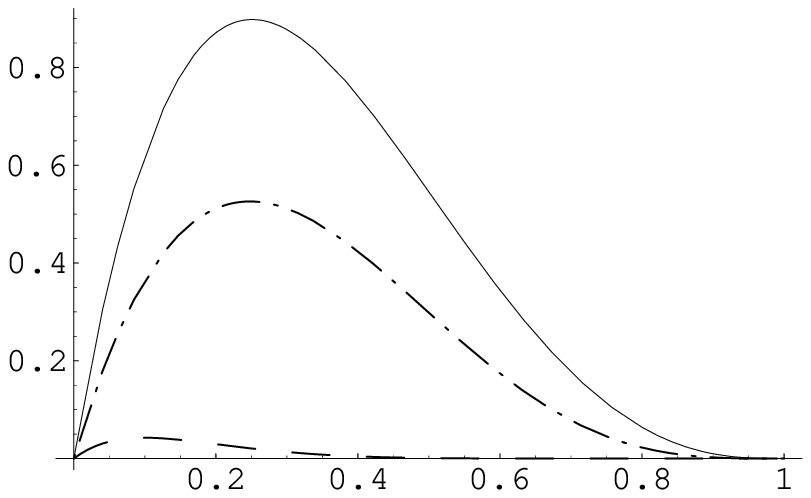,width=8.5cm,height=6.5cm}}
 \caption{The variation of valence quark distributions $x u_{{\rm val}}$ (Solid line), $x d_{{\rm val}}$ (Dot-Dashed) and $x s_{{\rm val}}$ (Dashed) with the
 Bjorken scaling variable $x$. \label{udsval}}
  \end{figure}

  \begin{figure}
  \centerline{\psfig{figure=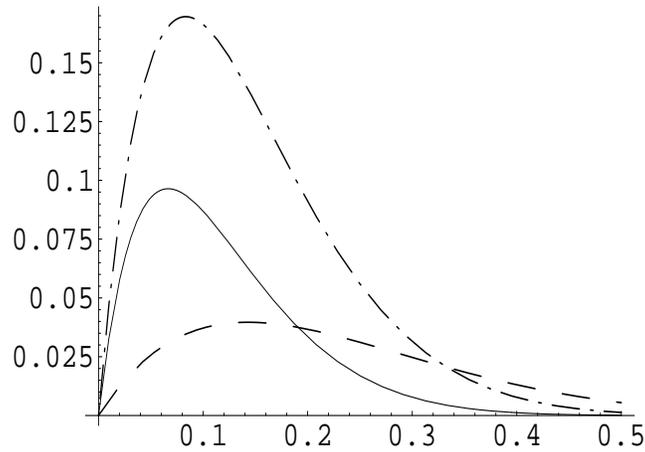,width=8.5cm,height=6.5cm}}
 \caption{The variation of sea quark distributions $x \bar u$ (Solid line), $x \bar d$ (Dot-Dashed) and $ x \bar s$ (Dashed) with the
 Bjorken scaling variable $x$. \label{udssea}}
  \end{figure}

\begin{figure}
  \centerline{\psfig{figure=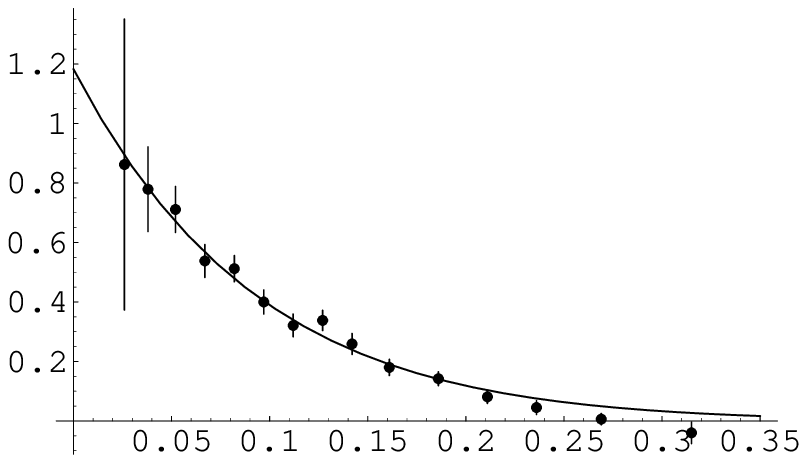,width=8.5cm,height=6.5cm}}
 \caption{The variation of $\bar d(x)-\bar u(x)$ asymmetry with the
 Bjorken scaling variable $x$ in comparison
 with the E866 data \cite{e866} at different $x$ values. \label{bdu}}
  \end{figure}

  \begin{figure}
  \centerline{\psfig{figure=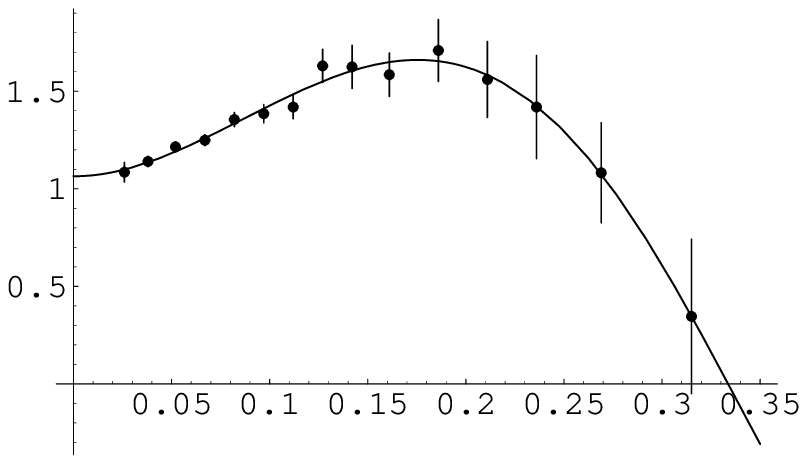,width=8.5cm,height=6.5cm}}
 \caption{The  variation of $\bar d(x)/\bar u(x)$ asymmetry  with the
 Bjorken scaling variable $x$ in comparison
 with the E866 data \cite{e866} at different $x$ values. \label{dbyu}}
  \end{figure}

 \begin{figure}
  \centerline{\psfig{figure=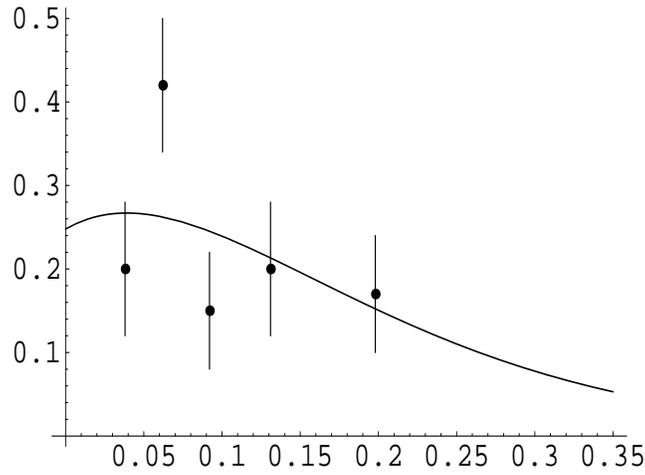,width=8.5cm,height=6.5cm}}
 \caption{The variation of $\frac{\bar d(x)-\bar u(x)}{u(x)-d(x)}$ asymmetry with the
 Bjorken scaling variable $x$ in comparison
 with the HERMES data \cite{hermes} at different $x$ values. \label{herm-asym}}
  \end{figure}

  \begin{figure}
  \centerline{\psfig{figure=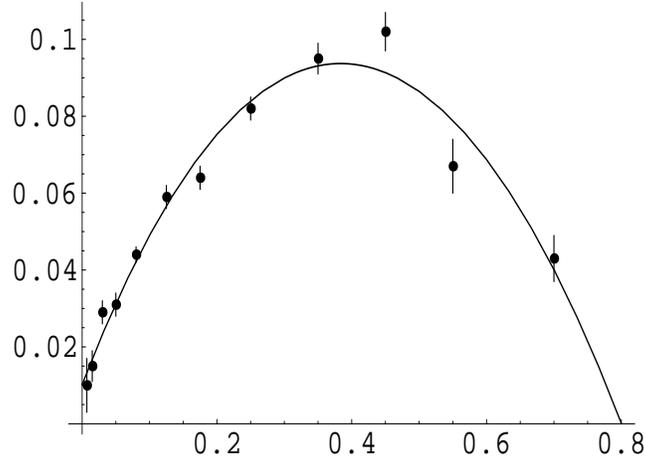,width=8.5cm,height=6.5cm}}
 \caption{The  variation of the difference of proton and neutron structure functions
 $F_2^p(x)-F_2^n(x)$ with the
 Bjorken scaling variable $x$ in comparison with the NMC data \cite{nmc-fpfn} at different $x$ values. \label{fpfn}}
  \end{figure}

  \begin{figure}
  \centerline{\psfig{figure=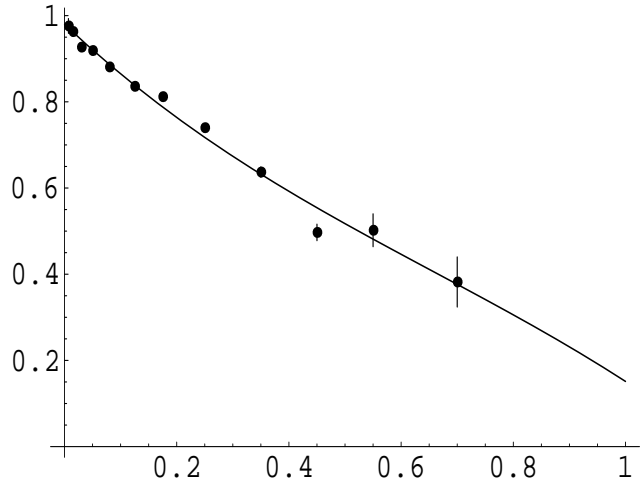,width=8.5cm,height=6.5cm}}
 \caption{The  variation of the ratio of neutron and proton structure functions
 $R^{np}(x)=\frac{F_2^n(x)}{F_2^p(x)}$ with the
 Bjorken scaling variable $x$ in comparison with the NMC data \cite{nmc-fpfn} at different $x$
values. \label{rnp}}
  \end{figure}

\begin{table}
\begin{center}
\begin{tabular}{|ccc|}\hline
Parameter & Data  & $\chi$CQM$_{{\rm config}}$
\\  \cline{3-3}
&  & $\zeta=-0.10$\\ & &    $a=0.13$\\ & &  $\alpha=0.45$ \\ & &
 $\beta=0.45$
\\
 \hline
$\Delta u^*$ & 0.85 $\pm$ 0.05 \cite{emc}  & 0.913
\\ $\Delta d$ & $-$0.41  $\pm$ 0.05 \cite{emc}   & $-$0.364 \\ $\Delta s$ &$-$0.07  $\pm$ 0.05
\cite{emc}  &$-$0.02 \\ $\Delta_3^*$ & 1.267 $\pm$ 0.0035
\cite{PDG}
 & 1.267\\ $\Delta_8$ & 0.58 $\pm$ 0.025
{\cite{PDG}} & 0.59 \\ $\Delta \Sigma$ & 0.19 $\pm$ 0.025
{\cite{PDG}}&  0.27
\\$\mu_{p}$ & 2.79$\pm$0.00 {\cite{PDG}}  & 2.81 \\
 $\mu_{n}$ & $-1.91\pm$0.00 {\cite{PDG}}&
 $-$1.96  \\ $\mu_{\Sigma^-}$ & $-1.16\pm$0.025
{\cite{PDG}}&  $-$1.19 \\ $\mu_{\Sigma^+}$ & 2.45$\pm$0.01
{\cite{PDG}}  & 2.46
\\ $\mu_{\Xi^o}$ & $-1.25\pm$0.014 {\cite{PDG}}  & $-$1.26 \\
$\mu_{\Xi^-}$ & $-0.65\pm$0.002 {\cite{PDG}}  & $-$0.64
\\
$\bar u-\bar d^*$ & $-0.118 \pm$ 0.015 \cite{e866}  & $-0.117$ \\

$\bar u/\bar d^*$ & 0.67 $\pm$ 0.06 {\cite{e866}}  & 0.67  \\

$I_G$ & 0.254  $\pm$ 0.005   & 0.255\\ $f_3$ &$-$ & 0.209\\

$f_8$  &$-$ & 1.06 \\

$f_3/f_8$ & 0.21 $\pm$ 0.05 {\cite{cheng}}  & 0.20 \\ \hline
{\small $*$ Input parameters} && \\ \hline
\end{tabular}
\end{center}
 \caption{The calculated values of the spin polarization
functions, baryon octet  magnetic moments and the quark flavor
distribution functions. The value of the mixing angle $\phi$ is
taken to be $20^o$. } \label{parameters}
\end{table}

\begin{table}

\begin{center}
\begin{tabular}{|c|ccc|ccc|} \hline
& \multicolumn{3}{c|} {$F_2^p(x)-F_2^n(x)$} & \multicolumn{3}{c|}
{$R^{np}(x)=\frac{F_2^n(x)}{F_2^p(x)}$}
\\ \cline{2-7}
$\langle x \rangle$ & Data& $\chi$CQM & $\chi$CQM$_{{\rm config}}$
&Data& $\chi$CQM & $\chi$CQM$_{{\rm config}}$
\\ \hline
0.007 & $0.010 \pm 0.007$ & 0.028 & 0.013 & $0.976 \pm 0.017$ &
1.294& 0.971  \\

0.015& $0.015 \pm 0.004$ & 0.054&  0.017 & $0.963 \pm
0.011$&1.241& 0.962\\

0.030& $0.029 \pm 0.003$& 0.053 & 0.023 & $0.927 \pm 0.007$& 1.258
&0.943\\

0.050& $0.031 \pm 0.003$& 0.063 & 0.031 & $0.919 \pm 0.007$& 1.228
&0.920\\

0.080& $0.044 \pm 0.002$& 0.073 & 0.042 & $0.881 \pm 0.006$& 1.217
&0.886\\

0.125& $0.059 \pm 0.003$& 0.085 & 0.057 & $0.836 \pm 0.007$& 1.170
&0.838\\

0.175& $0.064 \pm 0.003$& 0.098 &0.069 & $0.812 \pm 0.009$&
1.120&0.788\\

0.250& $0.082 \pm 0.003$& 0.106& 0.084 & $0.740 \pm 0.008$&
1.045&0.717\\

0.350& $0.095 \pm 0.004$& 0.112& 0.093 & $0.637 \pm 0.012$&
0.955&0.632\\

0.450& $0.102 \pm 0.005$ &0.110& 0.091 & $0.497 \pm 0.019$&
0.884&0.554\\

0.550& $0.067 \pm 0.007$& 0.102& 0.078 & $0.502 \pm 0.038$&
0.814&0.481\\

0.700& $0.043 \pm 0.006$& 0.058& 0.040 & $0.382 \pm 0.058$&
0.712&0.376\\

0.800& $-$& 0.046& 0.022 & $-$& 0.626 & 0.293\\

0.900 & $-$& 0.033 &0.016 & $-$& 0.602 &0.259\\

1& $-$ & 0.021&0.008& $-$&0.323& 0.239\\ \hline
\end{tabular}
\end{center}
\caption{The calculated values of the $F_2^p(x)-F_2^n(x)$ and
$\frac{F_2^n(x)}{F_2^p(x)}$ in $\chi$CQM and $\chi$CQM$_{{\rm
config}}$ at different values of $x$. }\label{fpfn-table}
\end{table}

\begin{table}

\begin{center}
\begin{tabular}{|cccc|} \hline
Quantity&   $x$ range & Data& $\chi$CQM$_{{\rm config}}$ \\ \hline
$\int{(\bar d(x)-\bar u(x))dx}$ &0-1& $0.118 \pm 0.012$ & 0.117
\\ $\int{(\bar d(x)-\bar u(x))dx}$ &0.05-0.35& $0.0803 \pm 0.011$ & 0.08\\
$\rho=2\frac{\int_0^1{x s(x)dx}}{\int_0^1{x(u(x)+d(x))dx}}$ &0-1&
$0.099^{+0.009}_{-0.006}$ & 0.09
\\ $\kappa=2\frac{\int_0^1{x s(x)dx}}{\int_0^1{x(\bar u(x)+\bar
d(x))dx}}$ &0-1& $0.477^{+0.063}_{-0.053}$& 0.47 \\
$\frac{\rho}{\kappa}= \frac{\int_0^1{x(\bar u(x)+\bar
d(x))dx}}{\int_0^1{x(u(x)+d(x))dx}}$ & 0-1 &0.2075 & 0.191 \\
\hline
\end{tabular}
\end{center}
 \caption{The calculated values of the $\bar d(x)-\bar u(x)$ asymmetry, $\bar d(x)-\bar u(x)$, $\rho$ and $\kappa$
 over a range of $x$.} \label{avg-table}
\end{table}


\begin{thebibliography}{99}

\bibitem{manohar}  S. Weinberg, Physica {\bf A 96}, 327 (1979);
A. Manohar and H. Georgi, Nucl. Phys. {\bf B 234}, 189 (1984).

\bibitem{emc} EMC Collaboration, J. Ashman {\it et al.},
Phys. Lett. {\bf  206B}, 364 (1988); Nucl. Phys. {\bf B 328}, 1
(1989); SMC Collaboration, B. Adeva {\it et al.}, Phys. Lett. {\bf
302B}, 533 (1993); P. Adams {\it et al.}, Phys. Rev. D {\bf 56},
5330 (1997); E142 Collaboration, P.L. Anthony {\it et al.},
 Phys. Rev. Lett. {\bf 71}, 959 (1993);
 E143 Collaboration, K. Abe {\it et al.},
 Phys. Rev. Lett. {\bf 75}, 391 (1995).

 \bibitem{nmc}  New Muon Collaboration, P. Amaudruz {\it et al.},
 Phys. Rev. Lett. {\bf 66}, 2712 (1991); M. Arneodo {\it et al.},
 Phys. Rev. {\bf D 50}, R1 (1994).

\bibitem{e866} E866/NuSea Collaboration, E.A. Hawker {\it et al.},
 Phys. Rev. Lett. {\bf 80}, 3715 (1998); J.C. Peng {\it et al.},
 Phys. Rev. {\bf D 58}, 092004 (1998); R. S. Towell {\it et al.},
{\it ibid.} {\bf 64}, 052002 (2001).

\bibitem{GSR}  K. Gottfried, Phys. Rev. Lett. {\bf 18}, 1174 (1967).

\bibitem{eichten}  E.J. Eichten, I. Hinchliffe and C. Quigg,
 Phys. Rev. {\bf D 45}, 2269 (1992).

\bibitem{cheng}  T.P. Cheng and Ling Fong Li, Phys. Rev. Lett.
{\bf 74}, 2872 (1995); hep-ph/9709293; Phys. Rev. {\bf D 57}, 344
(1998).

\bibitem{song} X. Song, J.S. McCarthy and H.J. Weber, Phys. Rev.
{\bf D 55}, 2624 (1997); X. Song, Phys. Rev. {\bf D 57}, 4114
(1998).

\bibitem{johan}  J. Linde, T. Ohlsson and Hakan Snellman, Phys. Rev.
{\bf D 57}, 452 (1998); T. Ohlsson and H. Snellman, Eur. Phys. J.,
{\bf C 7}, 501 (1999).

\bibitem{hd} H. Dahiya and M. Gupta, Phys. Rev. {\bf D 64}, 014013 (2001);
{\it ibid.} {\bf D 66}, 051501(R) (2002); {\it ibid.} {\bf D 67},
074001 (2003); {\it ibid.} {\bf 67}, 114015 (2003).

\bibitem{baldit}  NA51 Collaboration, A. Baldit {\it et al.},  Phys. Lett.
{\bf 253B}, 252 (1994).

\bibitem{hermes} HERMES Collaboration, K. Ackerstaff  {\it et al.}, Phys. Rev. Lett.
{\bf 81}, 5519 (1998).

\bibitem{emc-fpfn} J.J. Aubert {\it et al.}, Nucl. Phys. {\bf B 293}, 740
(1987).

\bibitem{nmc-fpfn} M. Arneodo {\it et al.}, Phys. Rev. {\bf D 50}, R1
(1994).

\bibitem{dgg} A. De Rujula, H. Georgi and S.L. Glashow,
Phys. Rev. {\bf D 12}, 147 (1975).

\bibitem{Isgur} N. Isgur, G. Karl and R. Koniuk, Phys. Rev. Lett.
{\bf 41}, 1269 (1978); N. Isgur and G. Karl,  Phys. Rev. {\bf D
21}, 3175 (1980); N. Isgur {\it et al.},  Phys. Rev.  {\bf D 35},
1665 (1987); P. Geiger and N. Isgur,  Phys. Rev. {\bf D 55}, 299
(1997); N. Isgur, Phys. Rev. {\bf D 59}, 034013 (1999).

\bibitem{yaouanc}  A. Le Yaouanc, L. Oliver, O. Pene and J.C. Raynal,
 Phys. Rev. {\bf D 12}, 2137 (1975);  {\it ibid.}. {\bf 15}, 844
 (1977).

\bibitem{mitra} M. Gupta and A.N. Mitra,
Phys. Rev. {\bf D 18}, 1585 (1978).


\bibitem{cdhs} CDHS Collaboration, H. Abramowicz {\it et al.}, Z. Phys. C17,283 (1983);
Costa {\it et al.}, Nucl. Phys. {B297}, 244 (1988).


\bibitem{reya} M. Gl{\rm $\ddot{u}$}ck, E. Reya, and A. Vogt, Z. Phys. {\bf C 67}, 433
(1995); M. Gl{\rm $\ddot{u}$}ck, E. Reya, M. Stratmann, and W.
Vogelsang, Phys. Rev. {\bf D 53}, 4775 (1996).

\bibitem{cqsm}  H. Weigel, Phys. Rev. {\bf D 55}, 6910 (1997);
M. Wakamatsu, Phys. Rev. {\bf D 67}, 034005 (2003).

\bibitem{bag} A.I. Signal and A.W. Thomas, Phys. Rev. {\bf D 40}, 2832 (1989).

\bibitem{alwall}  J. Alwall and G. Ingelman, Phys. Rev. {\bf D 71}, 094015 (2005).

\bibitem{stat} L.A. Trevisan, T. Frederico and L. Tomio, Eur.
Phys. J. {\bf C} (1999).

\bibitem{full} P.N. Pandit, M.P. Khanna and M. Gupta,
J. Phys. G {\bf 11}, 683 (1985).

\bibitem{mgupta1} M. Gupta and N. Kaur,
Phys. Rev. {\bf D 28}, 534 (1983);  M. Gupta, J. Phys. G: Nucl.
Phys. {\bf 16}, L213 (1990).


\bibitem{yaoubook} A. Le Yaouanc {\it et al.},
{\it Hadron Transitions in the Quark Model}, Gordon and Breach,
1988.

\bibitem{ao} J. Grasser, H. Leutwyler and M.E. Saino, Phys. Lett. {\bf  253B},
252 (1991); A.O. Bazarko {\it et al.}, Z. Phys {\bf C 65}, 189
(1995);  S.J. Dong, J.F. Lagae, K.F. Liu, Phys. Rev. Lett. {\bf
75}, 2096 (1995);  M. Goncharov {\it et al.}, Phys. Rev. {\bf D
64}  112006 (2001).

\bibitem{bjor} J.D. Bjorken, Phys. Rev. {\bf 148}, 1467 (1966); Phys. Rev. {\bf
D 1}, 1376 (1970).

\bibitem{ellis}  J. Ellis and R.L. Jaffe, Phys. Rev.  {\bf D 9},
1444 (1974); {\it ibid.} {\bf 10}, 1669 (1974).


\bibitem{neu charge}
 M. Gupta and A.N. Mitra, Phys. Rev. {\bf D 18}, 1585 (1978);
N. Isgur, G. Karl and D.W.L. Sprung,  {\it ibid} {\bf 23}, 163
(1981).

\bibitem{zeta} H. Dahiya, M. Gupta and J.M.S. Rana,
Int. Jol. of Mod. Phys. A, Vol. 21, No. 21, 4255 (2006).

\bibitem{PDG}  W.-M.Yao {\it et al.},
J. Phys. G  {\bf 33}, 1 (2006).


\end{thebibliography}
\end{document}